\documentclass[preprint]{aastex}
\usepackage{graphicx}
\usepackage{color}

\newcommand{\thisevent}{OGLE-2009-BLG-020}
\newcommand{\template}{HIP63762}

\newcommand{\RV}{{\rm RV}}
\newcommand{\lag}{{\rm lag}}
\newcommand{\bdv}[1]{\mbox{\boldmath$#1$}}

\def\bpi{{\bdv\pi}}
\def\btheta{{\bdv\theta}}
\def\bmu{{\bdv\mu}}
\def\bgamma{\bdv\gamma}
\def\e{{\rm E}}
\def\rel{{\rm rel}}
\def\geo{{\rm geo}}
\def\bs{{\bf s}}
\def\au{{\rm AU}}
\def\br{{\bf r}}
\def\bv{{\bf v}}

\shorttitle{RVs and Microlensing}
\shortauthors{Yee et al.}

\begin{document}

\title{Two Stars Two Ways:  
 Confirming a Microlensing Binary
Lens Solution with a Spectroscopic Measurement of the Orbit}

\author{
Jennifer C. Yee\altaffilmark{1,2},
John Asher Johnson\altaffilmark{1},
Jan Skowron\altaffilmark{3},
Andrew Gould\altaffilmark{4},
J. Sebastian Pineda\altaffilmark{5},
Jason Eastman\altaffilmark{1},
Andrew Vanderburg\altaffilmark{1},
Andrew Howard\altaffilmark{6}
}

\altaffiltext{1}{Harvard-Smithsonian Center for Astrophysics, 60 Garden Street, Cambridge, MA 02138 USA; jyee,jjohnson,jason.eastman,avanderburg@cfa.harvard.edu}
\altaffiltext{2}{Sagan Fellow}
\altaffiltext{3}{Warsaw University Observatory, Al. Ujazdowskie 4, 00-478 Warszawa, Poland}
\altaffiltext{4}{Department of Astronomy, Ohio State University, 140 West 18th Avenue, Columbus, OH 43210, USA}
\altaffiltext{5}{Department of Astronomy, California Institute of Technology, 1200 East California Boulevard, MC 249-17, Pasadena, CA 91125, USA}
\altaffiltext{6}{Institute for Astronomy, University of Hawaii, 2680 Woodlawn Drive, Honolulu, HI 96822-1839, USA}

\begin{abstract}
Light curves of microlensing events involving stellar binaries and planetary systems can provide information about the orbital elements of the system due to orbital modulations of the caustic structure. Accurately measuring the orbit in either the stellar or planetary case requires detailed modeling of subtle deviations in the light curve. At the same time, the natural, Cartesian parameterization of a microlensing binary is partially degenerate with the microlens parallax. Hence,
it is desirable to perform independent tests of the predictions of microlens orbit models using radial velocity time series of the lens binary system. To this end, we present 3.5 years of RV monitoring of the binary lens system \thisevent L, for which \citet{Skowron11} constrained all internal parameters of the 200--700 day orbit. Our RV measurements reveal an orbit that is consistent with the predictions of the microlens light curve analysis, thereby providing the first confirmation of orbital elements inferred from microlensing events. 
\end{abstract}


\section{Introduction}

While the general--relativistic microlensing effect has been
repeatedly observed, very few direct tests of the microlensing model
solutions have been possible. This is because microlensing is
inherently a rare transient phenomenon, and the lenses observed are
often extremely faint. Because a microlensing event requires that two
stars at different distances align with each other along the
line-of-sight to better than $\sim 1$ mas, in the densest parts of the
sky only about 1 in a million stars is expected to be
undergoing a microlensing event at any given time
\citep{Paczynski91, HanGould95}. These events are relatively brief ($\sim 2$
months) and (effectively) unrepeatable. In addition, since the only
light required to study the gravitational potential of the foreground
lens is provided by the background source, the lens itself can, in
principle, be completely non-luminous. Typically, lens stars are M
dwarfs that reside more than halfway to the center of the Galaxy at $\gtrsim
4$~kpc, and are thus very faint, making them extremely difficult or
impossible to followup after the event is over. This means that while
the ensemble of microlensing detections is robust, very few individual
lens stars can be studied in detail aside from what can be learned
from the lensing event. More importantly, there have been relatively
few confirmations of the complex microlensing modeling process even as
the number of parameters expands to include more effects.

Most previous tests of microlensing models have focused on confirming
that the measured brightness of the lens star is consistent with the
predicted lens mass in microlensing.  \citet{Bennett10_109} made an
independent confirmation of a microlensing model based on adaptive
optics observations of OGLE-2006-BLG-109. They demonstrated that the
observed lens flux is consistent with the predicted lens mass and
distance made by measuring the parallax and the finite source effects
in the light curve. Likewise \citet{Pietrukowiczetal05} discovered a
transient in M22 and classified it as likely microlensing event caused
by the lens belonging M22 magnifying the background bulge star. They
gave mass estimation based on the measured Einstein time-scale, the
known distances and globular cluster proper motion ($M =
0.14^{+0.10}_{-0.02} M_{\odot}$), which was later confirmed by
\citet{Pietrukowiczetal12} using VLT/NACO to show that the measured
lens flux corresponds to a mass of $M=0.18 M_{\odot}$. In addition,
\citet{Gould04} confirmed the microlens parallax measured for
MACHO-LMC-5 by showing that the direction of the relative proper
motion measured from the separation of the source and lens by
\citet{Alcock01} was consistent with the expectation from the
parallax.

In this paper, we present the first direct test of a microlensing
detection of orbital motion. While the orbital period at a typical Einstein
ring radius of a few AU (where microlensing sensitivity to companions
is maximized) is generally much longer than the typical lensing timescale of $t_{\rm E} \sim 20$ days, in some cases it is still possible to observe Keplerian motion of a binary lens system. The theory of orbiting binary lenses was first explored by \citet{Dominik98}. In practice, the detectability of orbital motion depends on having features in the light curve that are well-separated
in time ($\sim 20\,$ days) and a relatively
short orbital period ($\lesssim$ hundreds of days). The orbital motion leads to
changes in the shape of the caustic structure, and hence the magnification pattern due to the changing separation between the components of the
lens and can also rotate the caustic on the plane of the sky.

MACHO 97-BLG-41 \citep{Alcock00} was the first binary
lens to show strong deviations from the assumption of a static binary,
but the initial interpretation was that the deviations were due to a
third body \citep[a circumbinary planet;][]{Bennett99}, and it was not until a
later analysis of an independent data set that the orbiting binary
solution was found \citep{Albrow00}. This controversy was definitively
resolved by \citet{Jung13}, who conducted a combined fit of the data and found
that direct comparison of the orbiting binary and circumbinary planet
models clearly preferred the binary. 
Orbital motion is also found in systems in which the companion is a
planet, and was in fact observed in the second planetary system
discovered by microlensing: OGLE-2005-BLG-071 \citep{Udalski05,Dong09_071}. Hence,
experience has demonstrated that it is important to take these effects
into account when fitting microlensing light curves.

At the same time, a test of an orbital motion model would greatly increase our confidence in the derived orbital parameters for both stellar and planetary microlenses. Introducing the orbital motion effect into the microlensing models by definition increases the number of free parameters, raising the concern that any improvement in the fit is due primarily to fitting systematics or correlations in the data. In addition, the orbital motion parameters are known to be correlated with other microlensing parameters and effects such as the orbital parallax effect \citep[the effect of the Earth's motion on the light curve; see ][]{Batista11} and xallarap (motion of the source due to a binary). Because of the correlation with parallax, a confirmation of the orbital motion solution will also translate into increased confidence in the measured mass from the parallax effect.

While tests of the orbital motion solutions for stellar lenses with
planetary companions will remain difficult even with 30m
class telescopes, it is possible that this test could be done for
binary star lenses because the orbital motion signal is so much larger
(e.g. radial velocities $\mathcal{O}$ km s$^{-1}$ for binaries rather
than $\mathcal{O}$ m s$^{-1}$ for planets).

One system seen to exhibit orbital motion is the binary star lens in
\thisevent, which was analyzed by \citet{Skowron11}. They found that
it was impossible to find a satisfactory fit to the microlensing light
curve without allowing for orbital motion of the lens. From this
analysis, they were able to place broad constraints on the Keplerian
parameters of the orbit (reproduced in the upper, right-hand panels of
Figure \ref{fig:jointconstraints}). In brief, the posteriors indicate a
$0.84 M_{\odot}$ primary with an M-dwarf companion in a 200--700 day
orbit with some amount of eccentricity.

What makes this system unusual is that it is exceptionally close for a
microlensing lens system, $D_{\rm L} \sim 1\,$ kpc, such that the lens
is clearly visible in the blended light. In fact, with an $I$ magnitude of 15.6, the lens is brighter
than the unmagnified source. Because the
system is so bright and the expected radial velocity (RV) signal from the
lens is so large ($\sim 10$ km s$^{-1}$), it is possible to confirm the orbital motion of the
lens system measured from the microlensing light curve with followup
observations.

In this paper, we present radial velocity followup observations of the
lens system and confirm a microlensing orbit solution for the first
time. We begin by comparing and contrasting the direct observables of binary stars as seen with radial velocity and microlensing (Section \ref{sec:orbitpar}). The microlensing and RV observations of \thisevent\, are presented in Section \ref{sec:obs}. A detailed discussion of the RV data is given in Section \ref{sec:rvs} including a novel method to use the source star as a wavelength reference (Section \ref{sec:keckrvs}) and the final RV solution to the orbit (Section \ref{sec:rvorbit}). In Section \ref{sec:independent}, we show that this RV solution is consistent with the constraints on the orbit from \citet{Skowron11}, and in Section \ref{sec:joint} we perform a joint fit to both the RV and microlensing data to find the final parameters of the binary system. Our conclusions are given in Section \ref{sec:conclude}.

{\section{Orbit Kinematics: RV vs. Microlensing
    Observables}\label{sec:orbitpar}}

\begin{deluxetable}{lll}
\tablecaption{Standard, Keplerian Parameterization of a Binary\label{tab:kepbinarypar}}
\tablehead{
\colhead{Variable}&\colhead{Units}&\colhead{Meaning}
}
\startdata
\sidehead{Binary Orbit Parameters:}
~~~$t_{\rm peri}$      & days        & Time of periastron\\
~~~$a$                 & AU          & Semi-major axis\\
~~~$e$                 & \nodata     & Eccentricity\\
~~~$\Omega_{\rm node}$ & deg         & Longitude of ascending node\\
~~~$i$                 & deg         & Inclination\\
~~~$\omega_{\rm peri}$ & deg         & Argument of periastron\\
~~~$m_1$               & $M_{\odot}$ & Mass of primary\\
~~~$m_2$               & $M_{\odot}$ & Mass of secondary\\
\sidehead{Phase-Space Parameters:}
~~~$D_L$     & kpc           & Distance to system\\
~~~$\btheta$ & deg           & Coordinates of the system (RA/DEC)\\
~~~$\bmu$    & mas yr$^{-1}$ & Proper motion of the system\\
~~~$v_z$     & km s$^{-1}$   & Radial velocity of the system\\
\enddata
\end{deluxetable}

\begin{deluxetable}{lll}
\tablecaption{RV Parameterization of a Binary\label{tab:rvbinarypar}}
\tablehead{
\colhead{Variable}&\colhead{Units}&\colhead{Meaning}
}
\startdata
\sidehead{RV Orbit Invariants:}
~~~$t_{\rm peri}$      & days        & Time of periastron\\
~~~$P$                 & days        & Period\\
~~~$e$                 & \nodata     & Eccentricity\\
~~~$\omega_{\rm peri}$ & deg         & Argument of periastron\\
~~~${\cal M} =(m_2\sin i)^3/(m_1+m_2)^2$ & & Mass function\\
~~~($m_1$               & $M_{\odot}$ & Spectroscopic mass of primary)\\
\sidehead{Other known/measured parameters:}
~~~$\btheta$ & deg           & Coordinates of the system (RA/DEC)\\
~~~$v_z$     & km s$^{-1}$   & Radial velocity of the system\\
\enddata
\end{deluxetable}

\begin{deluxetable}{lll}
\tablecaption{Microlensing Parameters\label{tab:ulenspar}}
\tablehead{
\colhead{~~~Variable}&\colhead{Units}&\colhead{Meaning}
}
\startdata
\sidehead{13 Parameters of a Microlensing Model:}
~~~$u_0$ & $\theta_\e$ & Closest approach between the source and lens\\
~~~$t_0$ & days & Time when $u(t)=u_0$\\
~~~$t_\e$& days & Einstein crossing time\\
~~~$\rho$& $\theta_\e$& Normalized source size\\
~~~$\bpi_\e$& \nodata & Microlens parallax vector\\
~~~$s$ & $\theta_\e$ & Projected separation of the lens components\\
~~~$q$ & \nodata & Mass ratio between the lens components\\
~~~$\alpha$ & rad & Angle between the binary axis and the source trajectory\\
~~~$\bgamma$ & \nodata & Normalized, projected relative velocity of the binary\\
~~~$s_z$ & $\theta_\e$ & Relative position of the lens companion along the line of sight\\
~~~$\gamma_z$ & $\theta_\e$ & Relative velocity of the binary along the line of sight\\
\hline
\sidehead{Additional Known Parameters:}
~~~$\btheta$ & deg & Angular coordinates of the microlensing event (RA/DEC)\\
~~~$\theta_{\star}$ & mas & Angular size of the source\\
~~~$\pi_{\rm S}$ & mas & Source parallax\\
\enddata
\end{deluxetable}

\begin{deluxetable}{llll}
\tablecaption{Parameters Derived from Microlensing Parameters\label{tab:binarypar}}
\tablehead{
\colhead{~~~Variable}&\colhead{Definition}&\colhead{Units}&\colhead{Meaning}
}
\startdata
\sidehead{Intermediate Parameters:}
~~~$\theta_\e$ & $\theta_{\star}/\rho$ & mas & Angular size of the Einstein ring\\
~~~$\mu_{\rm geo}$ & $\theta_\e/t_\e$ & mas yr$^{-1}$ & Geocentric relative
  proper motion between the source \\
  & & & and lens (magnitude)\\
~~~$\hat{\bmu}_{\rm geo}$ & $\hat{\bpi}_\e$ & \nodata & Direction of
  geocentric relative proper motion\\
~~~$\bs$ & ($s\cos\alpha, s\sin\alpha$) & $\theta_\e$ & Projected binary
  separation vector\\
~~~$\Delta\btheta$ & $\bs\theta_\e$ & mas & Angular binary separation\\
~~~$M_{\rm tot}$ & $\theta_\e/(\kappa\pi_\e)$& $M_{\odot}$ & Total mass of binary\\
\hline
\sidehead{Binary Parameters:}
~~~$m_1$ & $M_{\rm tot}/(1+q^{-1})$ & $M_{\odot}$ & Mass of primary primary\\
~~~$m_2$ & $m_1/q$ & $M_{\odot}$ & Mass of secondary\\
~~~$D_L$ & AU$/(\theta_\e\pi_\e + \pi_{\rm S})$ & kpc & Distance to the
  binary system\\
~~~$\br_{\perp}$ & $D_L\btheta$ & AU & Physical position of the binary system\\
~~~$\bv_{\perp}$ & $D_L\bmu_{\rm geo}$ & km s$^{-1}$ & Projected velocity
  of the binary system\\
~~~$\Delta\br_{\perp}$ & $D_L\Delta\btheta$ & AU & Physical projected
  separation of the secondary\\
~~~$\Delta r_z$ & $D_L s_z\theta_\e$ & AU & Relative position of the secondary along
  the line of sight\tablenotemark{a}\\
~~~$\Delta\bv_{\perp}$ & $D_L s \theta_\e\bgamma$& km s$^{-1}$ &
  Relative, projected velocity of the secondary\\
~~~$\Delta v_z$ & $D_L \gamma_z \theta_\e$ & km s$^{-1}$& Relative velocity of the
  secondary along the line of sight\tablenotemark{a}\\
~~~($v_z$ & \nodata & \nodata & {\it Unknown} systemic radial velocity)\\
\enddata
\tablenotetext{a}{Note that in microlensing there is a perfect
  degeneracy between solutions into and out of the plane of the sky,
  i.e. $(\Delta r_z, \Delta v_z)\rightarrow(-\Delta r_z, -\Delta
  v_z)$.}
\end{deluxetable}

Fourteen parameters are required to completely characterize the
kinematics of a binary star orbit.  These could be the six phase-space
coordinates of each body at a given time plus its mass, or any
non-degenerate set of combinations of these quantities, for example those given in Table \ref{tab:kepbinarypar}.  Single-line
spectroscopic (RV) observations yield measurements nine of these 14
parameters, while microlensing observations can potentially yield 13.
However, the natural parameterizations of these two characterizations
are very different.  Hence, before comparing microlensing predictions
with RV measurements, it is essential to understand each
parameterization.

\subsection{Radial Velocity Parameters}

RV observations are normally thought of as yielding five (out of
eight) parameters that characterize the internal motion of the binary,
plus a spectroscopic estimate of the mass of the primary.  These are the period $P$, the eccentricity $e$, the argument of
periastron $\omega$, the time of periastron $t_{\rm peri}$, and the
mass function ${\cal M} =(m_2\sin i)^3/(m_1+m_2)^2$. These are the same as the RV observables except that the invariant ${\cal M}$ is replaced by the observable $K$, the RV semi-amplitude.
Of the remaining parameters that are not measured, the longitude of ascending node
($\Omega$), is rarely of physical interest and is therefore usually
ignored.  Thus, there are only two parameters of interest that
are not measured, the inclination $i$ and the mass ratio $q =
m_2/m_1$.  Note that for double-lined binaries, $q$ is measured, while
for extreme mass ratios (e.g., planets), ${\cal M} \simeq (q\sin i)^3 m_1$.

For RV, three center-of-mass parameters are also known, namely the
measured system radial velocity, $v_z$, and the two-dimensional
position on the sky ($\btheta$, i.e. the coordinates of the system in right
ascension and declination).  The last two are so intrinsic to the
process of measurement that they are not normally even considered as
``measurements''.

Table \ref{tab:rvbinarypar} summarizes the binary parameters known from RV observations.

{\subsection{Microlensing Parameters}\label{sec:ulenspar}}

Since microlensing can in principle measure 13 parameters, the
simplest way to characterize these is to specify the parameter that
cannot be measured: the system radial velocity.  In addition,
microlensing cannot distinguish between $(\Omega, \omega_{\rm peri})$ and $(180\, {\rm deg}-\Omega, \omega_{\rm peri}-180\, {\rm deg}$ \citep{Skowron11}.  That is, the
parameters that can be derived from microlensing are identical to
those from astrometric measurements for similar reasons: namely that
microlensing effects derive from the time evolution of the projected
positions of the two components on the plane of the sky. 

However, from the standpoint of understanding the information content
of microlens binary solutions, the above description is a bit too
simple.  First, several of the microlensing parameters are quite
unfamiliar combinations of phase-space and masses.  More important,
the precision with which these parameters can be measured is highly
variable, with some parameters measured to fractions of a percent and
others usually measured only to within a factor of a few \citep[with notable exceptions in ][]{Shin11, Shin12}. Properly
understanding how RV and microlensing can be compared requires taking
these differences into account.

First we consider the parameters that are required to characterize a
caustic-crossing binary microlensing event, which can then be
transformed into 13 physical parameters of the binary. Seven
parameters describe the basic microlensing event. Three of these
describe the underlying point-lens event, i.e., the time of maximum
$t_0$, the impact parameter $u_0$ (in units of the angular Einstein
radius $\theta_\e$), and the Einstein timescale $t_\e$.  There are
three basic binary lens parameters, the mass ratio $q$ and the vector
projected separation $\bs$ of the companion (in units of $\theta_\e$)
relative to the direction of lens-source relative proper motion in the
geocentric frame $\bmu_\geo$.  This vector is frequently expressed as
$\bs = (s\cos\alpha,s\sin\alpha)$, where $\alpha$ is the angle between
the binary axis and the source trajectory.  Finally, there is the
normalized source size $\rho = \theta_*/\theta_\e$, which is required
to described the extended duration of the caustic crossing due to the
finite angular size of the source $\theta_*$. Of these 7 parameters
$(t_0,u_0,t_\e,q.\,\bs,\rho)$ all but two ($t_0,u_0$) enter the
13-parameter binary characterization.

The accelerated motion of Earth can induce sufficient distortions
in the lightcurve to measure the ``microlens parallax''  \citep{Gould92a}.
\begin{equation}
\bpi_\e \equiv {\pi_\rel\over \theta_\e}{\bmu_\geo\over\mu_\geo},
\end{equation} 
where $\pi_\rel = \au(D_L^{-1} - D_S^{-1})$ is the lens-source
relative parallax (see \citet{GouldHorne13} for a didactic explanation).

Because microlensing observations are short compared to an
orbital period, the orbital parameters are naturally formulated in
terms of Cartesian phase-space coordinates, rather than Kepler
invariants (as for RV). In addition, because the microlensing event is
entirely governed by the projected motion of the binary, the most
robustly measured parameter of the motion is the projected, relative
velocity of the binary, $\Delta\bv_{\perp}$. This is parameterized
as instantaneous rates of change of $s$ and $\alpha$ respectively,
which yield $\bgamma = (\gamma_\parallel,\gamma_\perp)\equiv
(ds/dt/s,d\alpha/dt)$.

The remaining two parameters\footnote{Note that as discussed in Appendix A of \citet{Skowron11}, the microlensing $z$ direction points toward the observer, i.e. $+z$ is a blueshift, which is opposite the convention for RV.} $(\Delta r_z,\Delta v_z)$ must be
inferred from the impact of acceleration on the second derivatives of
$s$ and $\alpha$. There is no particular reason to express the final
two parameters as $(\Delta r_z,\Delta v_z)$; they might, for example,
be just as well written as the instantaneous angular
acceleration. Regardless, because microlensing events last only a
small fraction of an orbital period, clearly any such measurements
must be substantially less precise than the other parameters.
Nevertheless, all 13 parameters have been well measured in at least 4
cases \citep[OGLE-2005-BLG-018, OGLE-2009-BLG-020,MOA-2011-BLG-090,
  OGLE-2011-BLG-0417;][]{Shin11,Skowron11,Shin12,Gould13_0417}.

We now explain how these light curve parameters can be transformed into
physical properties. With the exception of $v_z$ and the sign of
$\Delta r_z$, all 13 binary parameters can be recovered with various
combinations of the known parameters. 

As in RV, $\btheta$ is automatically measured. In addition, the
measurement of $\rho$ enables a determination of $\theta_\e$.  This is
because the fit to the light curve yields the source flux and so, if
there are measurements in two bands, the position of the source on an
instrumental color-magnitude diagram and hence the dereddened flux $F$
and surface brightness $S$, and so finally $\theta_* = \sqrt{F/\pi S}$
\citep{Yoo04}.  Then $\theta_\e = \theta_*/\rho$, which
allows the transformation of three others into more familiar form
\begin{equation}
\mu_\geo = {\theta_\e\over t_\e};
\qquad
\Delta\btheta = \bs\theta_\e,
\end{equation} 
where $\Delta\btheta$ is the instantaneous angular separation
between the two components. 

The combination of $\bpi_\e$ and $\theta_\e$ then add three more
binary parameters.  First, of course, $\bpi_\e$ immediately yields the
direction of $\bmu_\geo$.  Then, because 
\begin{equation}
\theta_\e^2 = \kappa M\pi_\rel,
\qquad \kappa \equiv {4 G\over \au\,c^2}\simeq 
8.1\,\frac{\rm mas}{M_\odot},
\end{equation}
\citep{Einstein36} and because $\pi_s$ is usually known quite well, we obtain the mass
and distance of the system
\begin{equation}
M = {\theta_\e\over \kappa \pi_\e};
\qquad
D_l = {\au\over \theta_\e\pi_\e + \pi_s}.
\end{equation}
The measurement of $D_L$ enables us to transform angular measurements
into physical measurements, i.e., $\Delta\br_\perp =
D_L\Delta\btheta$, $\bv_\perp = D_L\bmu_\geo$, and $\br_\perp =
D_L\btheta$.  Hence, we now have 9 parameters,
$(m_1,m_2,\br,\bv_\perp,\Delta\br_\perp)$ where $\br$ is the 3-space
position of the lens system at $t_0$.  Finally, via the other
measurement parameters, these yield $\Delta\bv_\perp=D_L
s\theta_\e\bgamma$ and so 13 parameters $(m_1$, $m_2$, $D_L$, $\br_\perp$, $\bv_\perp$, $\Delta\br_\perp$, $\Delta\bv_\perp$, $ \pm\Delta r_z$, $\Delta v_z)$.

Table \ref{tab:ulenspar} summarizes the microlensing parameters, and Table \ref{tab:binarypar} gives an overview of how those parameters translate into the parameters of a binary. Appendix B of \citet{Skowron11} provides additional details on the transformation between microlensing parameters and the parameters of a Keplerian orbit.

\subsection{RV vs. Microlensing: Points of Comparison}

To understand the conditions under which binary microlensing
observations can be tested by RV, we now focus on 10 parameters,
namely the six phase-space coordinates of internal motion
($\Delta\br,\Delta\bv$), the two masses (($m_1,m_2$), or equivalently
$(M,q)$), the system distance ($D_L$ or $r_z$) and the direction of
transverse motion ($\bmu/\mu$).  We therefore ignore the system radial
velocity ($v_z$), which can be measured very well by RV but not at all
by microlensing, the magnitude of the proper motion ($\mu$), for which
the reverse holds true, as well as the system angular coordinates
$(\btheta)$, which were included only for completeness.

RV observations measure 5 combinations of these quantities, namely
four internal phase-space coordinates and the mass function, which is
a combination of $m_1$, $m_2$, and $i$.  Here, we ignore for the
moment the fact that spectroscopic measurements also return an
estimate of the primary mass $m_1$.

We begin the analysis of microlensing by examining the ``typical good
case'', in which ($q$, $\bs$, $\theta_\e$, $\bpi_\e$, $\bgamma$) are measured,
while $(\Delta r_z,\Delta v_z)$ are not measured.  These eight
measured quantities are all combinations of the ten parameters under
consideration.  That is $(q,\bpi_\e,\theta_\e)\leftrightarrow
(m_1,m_2,D_L,\bmu/\mu)$, and $(\Delta \br_\perp,\Delta
\bv_\perp)\leftrightarrow D_L\theta_\e(\bs,s\bgamma)$.  Hence, with a
total of 8 microlensing plus 5 RV $=13$ constraints on a total of 10
parameters, there are nominally three independent points of comparison.

However, in the case of OGLE-2009-BLG-020, the situation is not quite so
favorable.  There is a well-known degeneracy between $\gamma_\perp$
and $\pi_{\e,\perp}$, the component of $\bpi_\e$ parallel to the
projection of Earth's acceleration on the plane of the sky \citep{Batista11}, and this degeneracy is present in the solution of
OGLE-2009-BLG-020 as well \citep{Skowron11}.  Hence, in fact, there
are only two independent points of comparison in the present
case. 

In addition, if we consider that RV is returning six of the 10
parameters (i.e., including the spectroscopic determination of the
mass) then the parameter counting yields $6+8-1-10=3$ independent
points of comparison.  In the present work, we will not consider the mass test
due to the low signal-to-noise of the spectra. Nevertheless, it is clear that with by combining microlensing and RV observations, the problem is over-constrained, allowing a direct test of the microlensing orbit prediction.

\section{Observations\label{sec:obs}}

\subsection{Microlensing Data}

The microlensing data on \thisevent\, are described in detail in Section 2 of \citet{Skowron11} and shown in their Figure 1. In brief, the event was monitored by the Optical Gravitational Lensing Experiment (OGLE) and Microlensing Observations in Astrophysics (MOA) survey groups. In addition, it was observed by eight followup telescopes: Bronberg Observatory 36cm, Campo Catino Austral Observatory 40cm, CTIO/SMARTS 1.3m, Farm Cove Observatory 36cm, Faulkes Telescope North  2.0m, Faulkes Telescope South 2.0m, Kumeu Observatory 36cm, and University of Tasmania 1.0m. As described in Section 2.2 of \citet{Skowron11}, the data have had outliers removed, been binned, and had the error bars rescaled so that $\chi^2_d/{\rm dof} \sim 1$ for each data set.

\subsection{Keck/HIRES}

Between Mar 2011 and Oct 2011 \thisevent\ was observed six times by the HIgh-Resolution Echelle Spectrometer \citep[HIRES; ][]{Vogt94} on the Keck I telescope with an exposure time of 700s. One additional observation was taken in Aug 2013 with an exposure time of 900s. All the spectra were taken without the iodine cell and $R\approx55,000$. A summary of observations is given in Table \ref{tab:rvtable}.

These data were reduced using the standard California Planet Search (CPS) pipeline \citep{Howard10}.

\subsection{Magellan/MIKE}

Seven of the observations were taken with the red camera of the Magellan Inamori Kyocera
Echelle spectrometer \citep[MIKE; ][]{Bernstein03} on the 6.5m Magellan/Clay telescope. The
observations are summarized in Table \ref{tab:rvtable}. Most of the
observations were taken with the $0.7^{\prime\prime}$ slit with a 900
second exposure time. The exceptions were observations on 24 May 2014
and 5 September 2014, which were taken in poor seeing conditions
($2^{\prime\prime}$ and $>3^{\prime\prime}$, respectively).

These spectra were reduced using the CarPy MIKE pipeline \citep{Kelson00,Kelson03}.
Each spectrum was reduced individually with the exception of the
observations of 5 September 2014. In that case, we use the MIKE
pipeline to stack the four spectra to increase the signal-to-noise ratio (S/N).
The wavelength calibration was done relative to ThAr lamp observations
taken before and/or after the science exposures. We removed the blaze by fitting the continuum of each order with a (Gaussian)+(Constant)+(Slope), i.e.,
\begin{equation}
\label{eqn:5pargauss}
f(x) = p_1\exp\left(\frac{-(x-p_2)^2}{2p_3}\right) + p_4 + p_5 x.
\end{equation}
This results in a flatter continuum than the standard CarPy reduction. However, the differences in the resultant RVs are minimal. We find that our method results in slightly
smaller RV uncertainties $\sim10$--20\%, but the difference in the
measured RVs between the two methods is ($\lesssim0.1$ km s$^{-1}$) is an order of magnitude smaller than the values of the uncertainties ($\sim 0.5$--$1$ km s$^{-1}$).

\section{Radial Velocities\label{sec:rvs}}

\begin{deluxetable}{rrrllrlrrr}
\rotate
\tablecaption{RVs of \thisevent}
\tablehead{
\colhead{BJD$^{\prime}$}&
\colhead{RV}&
\colhead{$\sigma_{\rm RV}$}&
\colhead{Observatory}&
\colhead{Instr}&
\colhead{Date\tablenotemark{a}}&
\colhead{Slit}&
\colhead{Exposure}&
\colhead{Resolution}&
\colhead{S/N @}\\
\colhead{} &
\colhead{(km s$^{-1}$)}&
\colhead{} &
\colhead{} &
\colhead{} &
\colhead{} &
\colhead{} &
\colhead{Time (s)} &
\colhead{} &
\colhead{6000\AA} 
\label{tab:rvtable}
}
\startdata
5635.1293 & -36.23 & 0.88 &         Keck I & HIRES & 14 Mar 2011 & 0.861" &  700 & 55,000 & 10\\
5668.0545 & -40.37 & 0.90 &         Keck I & HIRES & 16 Apr 2011 & 0.861" &  700 & 55,000 & 12\\
5708.0989 & -42.77 & 1.35 &         Keck I & HIRES & 26 May 2011 & 0.861" &  700 & 55,000 &  8\\
5723.9147 & -44.12 & 0.93 &         Keck I & HIRES & 11 Jun 2011 & 0.861" &  700 & 55,000 & 10\\
5797.7612 & -43.82 & 0.74 &         Keck I & HIRES & 24 Aug 2011 & 0.861" &  700 & 55,000 & 13\\
5843.7469 & -37.06 & 0.27 &         Keck I & HIRES &  9 Oct 2011 & 0.861" &  700 & 55,000 &  8\\
6530.7713 & -42.96 & 1.63 &         Keck I & HIRES & 26 Aug 2011 & 0.861" &  900 & 55,000 & 12\\
6586.4889 & -44.38 & 0.85 &  Magellan/Clay & MIKE  & 20 Oct 2013 & 0.7"   &  900 & 31,000 & 10\\
6586.5058 & -44.30 & 0.52 &  Magellan/Clay & MIKE  & 20 Oct 2013 & 0.7"   &  900 & 31,000 & 12\\
6721.8966 & -33.81 & 0.99 &  Magellan/Clay & MIKE  &  4 Mar 2014 & 0.7"   &  900 & 31,000 & 15\\
6722.8966 & -33.23 & 0.56 &  Magellan/Clay & MIKE  &  5 Mar 2014 & 0.7"   &  900 & 31,000 & 17\\
6801.8221 & -42.13 & 0.98 &  Magellan/Clay & MIKE  & 24 May 2014 & 1.0"   &  900 & 22,000 &  7\\
6906.5487\tablenotemark{b} & -42.83 & 0.67 &  Magellan/Clay & MIKE  &  5 Sep 2014& 0.7"& 6900 & 31,000&  7\\
6908.6027 & -43.25 & 0.44 &  Magellan/Clay & MIKE  &  7 Sep 2014 & 0.7"   & 1800 & 31,000 & 25
\enddata
\tablenotetext{a}{Start of night}
\tablenotetext{b}{This observation is the sum of 4 exposures
  ($3\times1800$s and $1\times1500$s). The quoted signal-to-noise is
  for the summed spectrum.}
\end{deluxetable}

\begin{figure}
\includegraphics[width=\textwidth]{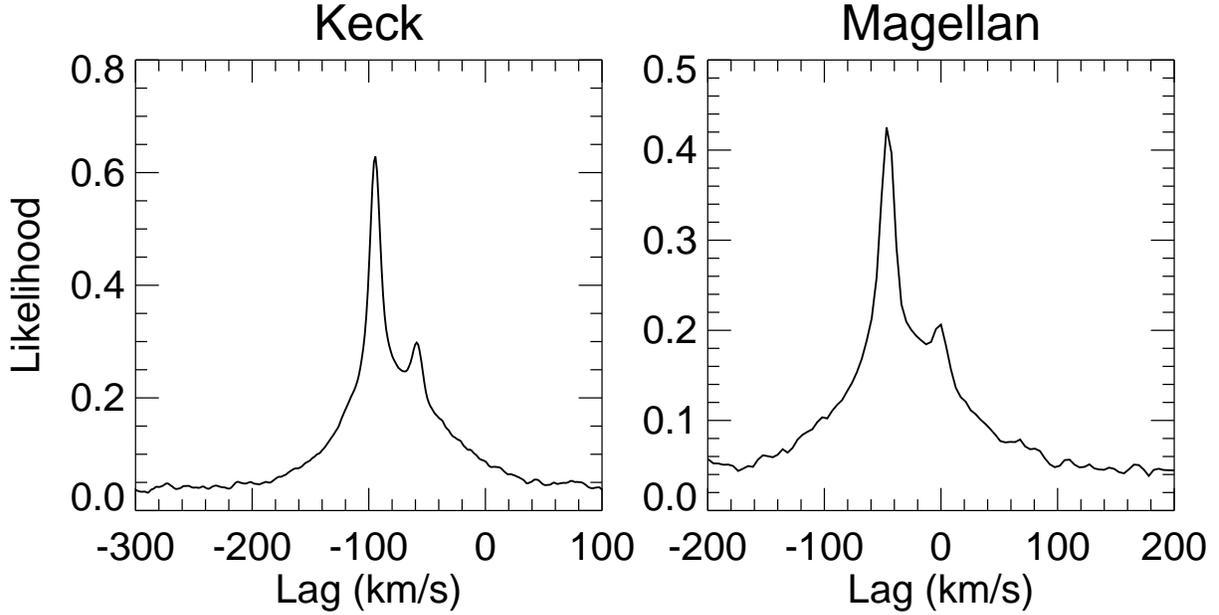}
\caption{Typical maximum likelihood functions for the spectrum of
  \thisevent\,from Keck/HIRES (left) and Magellan/MIKE (right). The
  taller peak corresponds to the lens primary and the fainter peak to
  the source star. The velocity scale is arbitrary. \label{fig:ccf}}
\end{figure}

\subsection{RV template}

To determine the radial velocities of the lens primary, we
cross-correlate the spectra of \thisevent\, with a template from the CPS
database. Because the default CPS pipeline only produces a flattened and
stitched spectrum of \template\, covering the wavelength range
$5010\AA$--$6309\AA$, we use only the orders of the HIRES spectrum of
\thisevent\,that overlap with that region (i.e. orders 0--14).
We combine the individual CCFs for each order to create a
maximum likelihood function (ML) following the prescription in
\citet{Zucker03}. This produces a double-peaked ML in which
the taller peak corresponds to the (brighter) lens and the smaller
peak to the source (see Figure \ref{fig:ccf}). 

We selected the template by cross-correlating the highest
S/N Keck spectrum (from BJD$^{\prime}=5797.7612$) with spectra of all stars in the CPS
database and measuring the heights of both the source and lens
peaks. We find that the lens is broadly consistent with being an early
K-dwarf and select our template from amongst the 10 best matches with
$5.5\leq M_V\leq8.5$ and $0.85\leq (B-V) \leq 1.35$ (cuts that encompass most stars with likelihood peaks within 95\% of the maximum value). Note that our template choice, \template, is also
broadly consistent with the source spectrum, obviating the need for a
two template fit. Because the source is a clump giant and the CPS sample specifically selects against giants, it is unsurprising that the best available template for the source would be a dwarf of similar color. 

\subsection{Keck Velocities\label{sec:keckrvs}}

To determine the radial
velocity of the lens and source for each epoch, we calculate the ML of \thisevent\, as compared to \template\, in the manner described above. We then fit the two tallest
peaks of the ML with 3, 4, and 5 parameter Gaussians (see Equation
\ref{eqn:5pargauss}), and take the result with the best $\chi^2$. The
measured lag of each peak is then the value of $p_2$ for the best
fit. 

Because the source star has a constant velocity (confirmed by applying a barycentric correction to the measured lag), we can use it as
a wavelength reference in a manner analogous to an iodine
cell. Both the source and the lens spectra encounter the same
systematic effects as their light passes through the telescope and
instrument optics. Hence, we take the measured RV of the lens to be
\begin{equation}
\RV_{\rm lens} = \lag_{\rm source} - \lag_{\rm lens},
\end{equation}
where ${\rm lag_{source}}$ and ${\rm lag_{lens}}$ are the lags of the
source and lens, respectively, measured relative to \template. This
automatically takes into account (and removes) the barycentric
velocity and other systematics induced by the instrument. The
uncertainty of the RV is taken to be the standard deviation of
the RVs measured for each individual order using the same method
(i.e., $\RV_{{\rm lens}, i} = \lag_{{\rm source},i} - \lag_{{\rm
    lens},i}$). The uncertainties in the RVs are dominated by uncertainties in the measurement of the source peak. These uncertainties are larger than they would be if we simply made a barycentric correction to the measured velocity of the lens peak. However, the RVs using our method are sufficiently precise for characterizing the system and likely to be more accurate because our method automatically accounts for systematic effects.
    The final velocities are given in Table
\ref{tab:rvtable} where BJD$^{\prime}={\rm BJD_{\rm TBD}}-2450000$.

\subsection{Magellan Velocities\label{sec:magrvs}}

For the Magellan/MIKE spectra, we used only the orders from the red camera that overlapped with
the iodine region, i.e. orders 57-68 (12 total). We also
cross-correlate those orders against the flattened Keck/HIRES spectrum of
\template. We create the ML and extract the lag of tallest peak in the
same way as for the Keck data. However, as can be seen from Figure
\ref{fig:ccf}, the ML (and CCFs) for the Magellan/MIKE spectra are lower resolution and lower S/N than the Keck/HIRES spectra, and so we are not able to reliably extract the lag of the
source using this method. Instead, we calculate the barycentric correction explicitly for each observation using {\tt BARYCORR} \citep{WrightEastman14} and apply it to the measured lens RV. This leaves a velocity offset between the Magellan RVs and the Keck RVs equivalent to the difference in radial velocity between the source star and the template.
We can place the Magellan lags on
the same velocity scale as the Keck RVs by calculating this difference, i.e.
the weighted mean lag of
the source star in the Keck data ($<\lag_{\rm source, K}>$) after
the barycentric velocity has been removed. Hence, the Magellan velocities on the
Keck system are given by
\begin{equation}
\RV_{\rm lens} = <\lag_{\rm source, Keck}> - (\lag_{\rm lens, Mag} - {\rm BC}).
\end{equation}
 To
compute the uncertainty in the radial velocities from the Magellan
data, we first compute the uncertainty in $\RV_{\rm lens}$ by computing the standard deviation of $\RV_{\rm lens}$
measured for each order individually, as we did for the Keck data. Then, we add this in quadrature to the standard deviation of $<\lag_{\rm
  source, Keck}>$ as measured from the Keck MLs ($\sigma_{\rm <lag_{source}>} = 0.29\,{\rm km\,s}^{-1}$). Final values
for the radial velocities are given in Table \ref{tab:rvtable}.

\subsection{RV Orbit\label{sec:rvorbit}}

\newcommand{\bjdtdb}{\ensuremath{\rm {BJD_{TDB}}}}
 \begin{deluxetable}{lcc}
\tablecaption{RV Parameters: Median values and 68\% confidence interval for \thisevent}
\tablehead{\colhead{~~~Parameter} & \colhead{Units} & \colhead{RV-only}\label{tab:rvpar}}
\startdata
\sidehead{RV Orbit Parameters:}
    ~~~$t_{\rm peri}$\dotfill &Time of periastron (\bjdtdb)\dotfill & $6142.7_{-1.9}^{+2.5}$\\
    ~~~$P$\dotfill &Period (days)\dotfill & $276.37_{-0.91}^{+0.96}$\\
     ~~~$e$\dotfill &Eccentricity\dotfill & $0.335_{-0.056}^{+0.074}$\\
    ~~~$\omega_{\rm peri}$\dotfill &Argument of periastron (degrees)\dotfill & $156.8\pm3.4$\\
    ~~~$\cal M$\dotfill & Mass Function\dotfill & $6.20_{-0.95}^{+2.08} \times 10^{-3}$\\
\hline
\sidehead{Other Parameters:}
    ~~~$a$\dotfill &Semi-major axis (AU)\dotfill & $0.8467\pm0.0076$\\
    ~~~$K$\dotfill &RV semi-amplitude (km/s)\dotfill & $6.370_{-0.450}^{+0.880}$\\
\enddata
\end{deluxetable}

\begin{figure}
\includegraphics[width=0.9\textwidth]{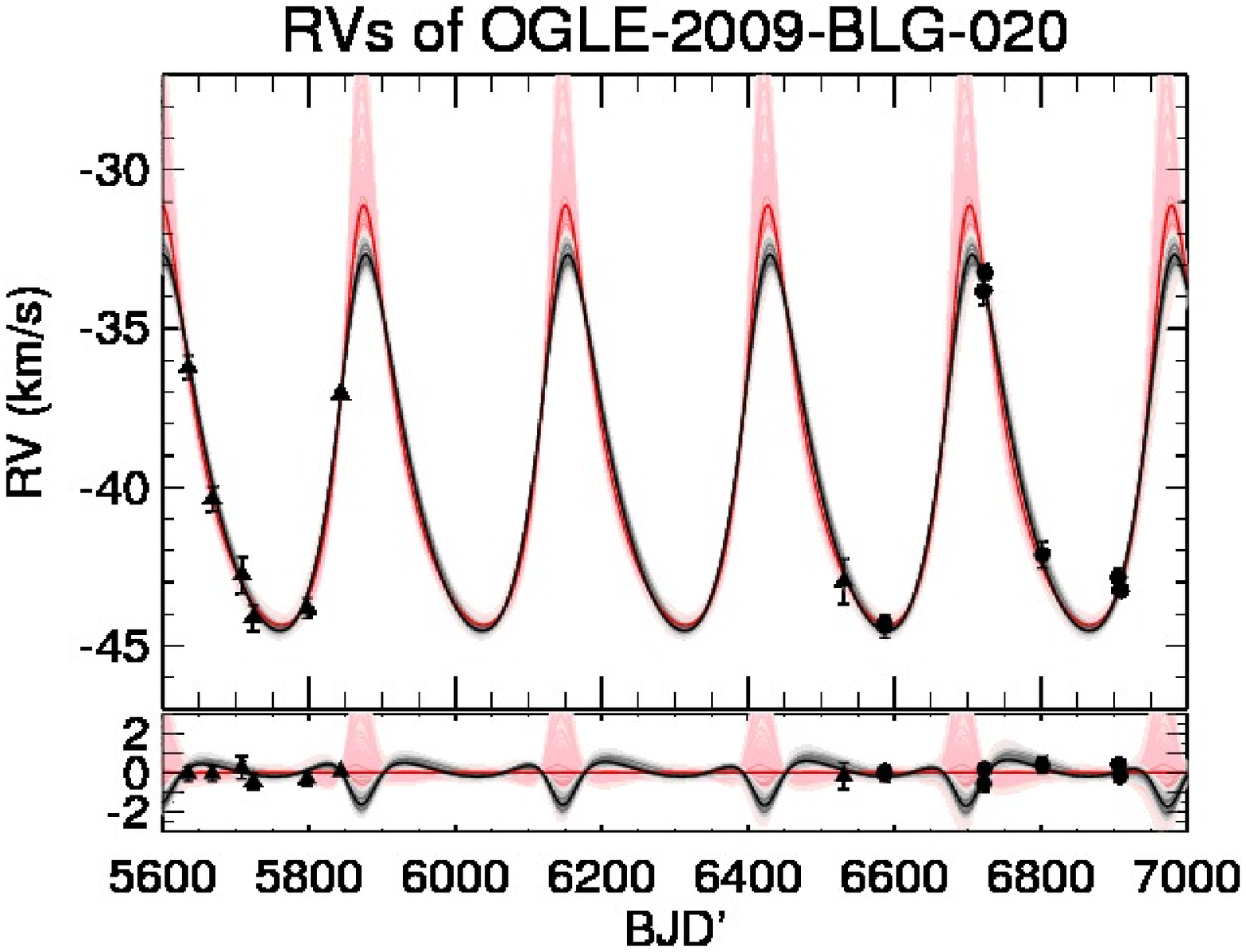}
\caption{Measured radial velocities of {\thisevent}L. The red line
  shows the best-fit orbit to the RV data alone, and the black line
  shows the best-fit orbit from the joint MCMC including both the RV
  and microlensing data. The gray (and pink) liness show joint (and RV-only) fits that are 1, 2,
  and 3-$\sigma$ from the best fit (reflected by the shading). \label{fig:rv}}
\end{figure}

We use {\tt EXOFAST} \citep{Eastman13} to find the best-fit orbit to the
radial velocity data of \thisevent. This package finds preliminary solutions using a Lomb-Scargle periodogram, refines them using an Amoeba minimization, and determines the uncertainties using a Markov Chain Monte Carlo. We began by finding a preliminary solution for the period using a Lomb-Scargle periodogram and then seed {\tt EXOFAST} with this value as a prior. We allow for free eccentricity but do not allow for a slope, such as might be caused by a third body in the system. {\tt EXOFAST} uses BJD as the time standard.

These fits clearly indicate that
our uncertainties for the radial velocities are over-estimated, and
rescales them by 0.4255. This fit shows that the lens has a period of $\sim 276$ days and an eccentricity of 0.341. The red line in Figure \ref{fig:rv} shows the best-fit RV curve to the data. The full RV solution (parameters and their uncertainties) is given in
Table \ref{tab:rvpar}. The posteriors are shown in the lower-left panels of Figure \ref{fig:sepconstraints}. Note that {\tt EXOFAST} provides the argument of periastron of the primary $\omega_{\star}$. We report the argument of periastron of the secondary $\omega_{\rm peri}\equiv \omega_{\star}-180\, {\rm deg}$. 

\section{Comparison of Independent Fits\label{sec:independent}}

\begin{deluxetable}{llrr}
\tablecaption{Joint Fit: Orbit Parameters}
\tablehead{
\colhead{Parameter}&\colhead{Units}&\colhead{Value}&\colhead{Uncertainty}
\label{tab:orbittable}
}
\startdata
Eccentricity         & \ldots       & $    0.265$ & $\pm  0.025$ \\
$a$                  & AU           & $    0.865$ & $\pm  0.024$ \\
$t_{\rm peri}$       & days         & $ 4758.692$ & $\pm  2.685$ \\
$\Omega_{\rm node}$  & deg          & $   -7.767$ & $\pm  1.260$ \\
inclination          & deg          & $  129.424$ & $\pm  1.238$ \\
$\omega_{\rm peri}$  & deg          & $  151.600$ & $\pm  3.325$ \\
$M_{\rm tot}$        & $M_{\odot}$  & $    1.132$ & $\pm  0.097$ \\
$D_{\rm L}$          & kpc          & $    0.747$ & $\pm  0.020$ \\
$q$                  & \ldots       & $    0.275$ & $\pm  0.003$ \\
\hline
Derived:\\
Period               & days         & $  276.555$ & $\pm  0.300$ \\
$m_1$                & $M_{\odot}$  & $    0.244$ & $\pm  0.021$ \\
$m_2$                & $M_{\odot}$  & $    0.888$ & $\pm  0.076$ \\
\enddata
\end{deluxetable}

\begin{figure}
\includegraphics[height=0.7\textheight]{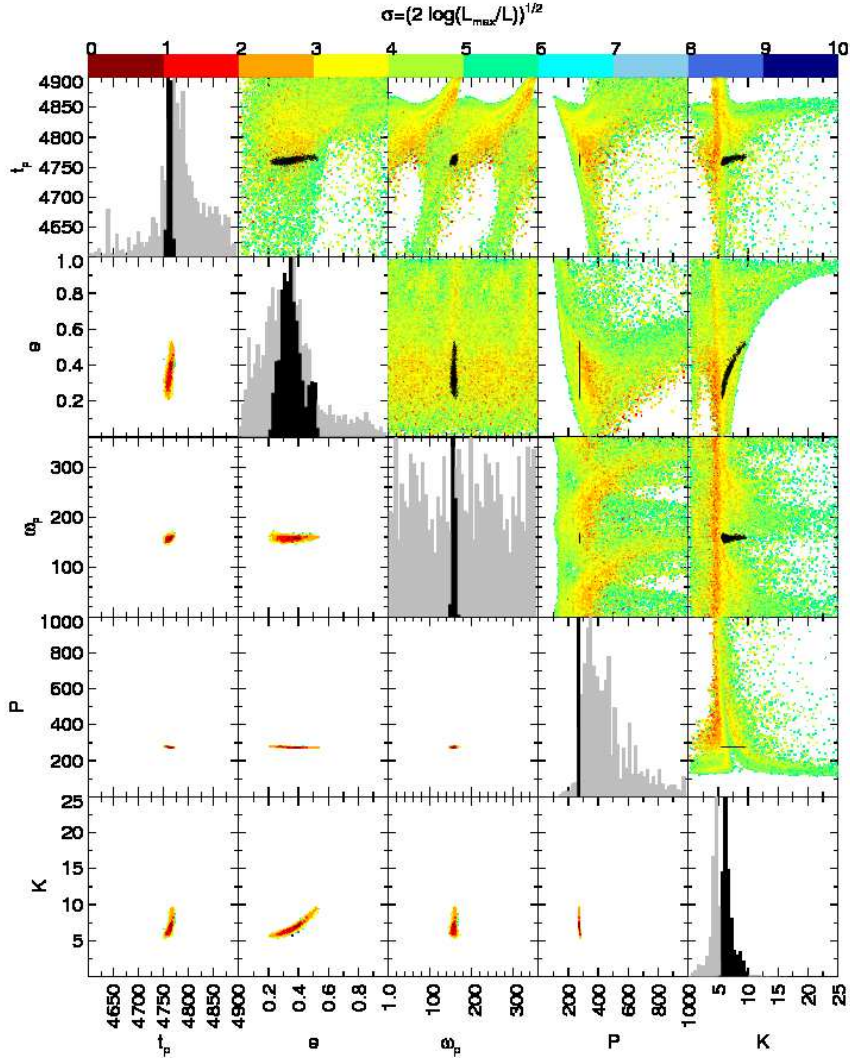}
\caption{A comparison of the orbit constraints for {\thisevent}L from
  independent fits to the microlensing light curve \citep[upper right; ][]{Skowron11} and the radial
  velocities (lower left; {\tt EXOFAST}, Section \ref{sec:rvorbit}). In the upper right panels, the black
  contours show the RV constraints overplotted on the
  microlensing constraints. The center panels on the diagonal compare
  the posteriors (gray: microlensing, black: radial velocity). For the microlensing constraints, the colors represent the weights of each link as described in \citet[Section 3.4 of ][]{Skowron11}. For the RVs, the colors reflect the likelihood as determined from the $\chi^2$ of each fit.
\label{fig:sepconstraints}}
\end{figure}

Figure \ref{fig:sepconstraints} compares the RV constraints on the
orbit of \thisevent L (Section \ref{sec:rvorbit}) derived from {\tt
  EXOFAST} to the independent constraints on the orbit from the
microlensing light curve \citep{Skowron11}. The microlensing
constraints include the weighting for the Jacobian, the Galactic
model, and lens flux as described in Sections 3.4 and 4.2.1 of
\citet{Skowron11}. For the purposes of this comparison, we extrapolate
$t_{\rm peri}$ for the RV fit backwards to the time of the
microlensing observations. In addition, note that microlensing uses
$HJD$ as the time standard rather than $BJD$, but this difference is
many orders of magnitude smaller than the uncertainties in the measured parameters.

The constraints on the RV parameters from the microlensing light curve are derived from the MCMC fits to the microlensing data in
\citet{Skowron11}.
Because of the $\pm$ degeneracy in $(s_z, \gamma_z)$ measured from microlensing,  there is a perfect degeneracy between $\omega_{\rm peri}$ and $\omega_{\rm peri}-180\,{\rm deg}$. Since both are equally valid, we plot both solutions in Figure \ref{fig:sepconstraints} leading to periodic behavior in $\omega_{\rm peri}$.

Figure \ref{fig:sepconstraints}œ clearly shows that the constraints on the orbit of
\thisevent L from the radial velocity data are consistent with the
observed properties of the orbit measured from the microlensing light
curve. This is the first confirmation of orbital motion of a 2-body
system measured from a microlensing light curve.

\section{Joint Fit\label{sec:joint}}

We also perform a joint MCMC fit to the RV and microlensing data to
determine the best constraints on the physical properties of the \thisevent L
system. The joint MCMC is performed in the same parameter space as in \citet{Skowron11}, i.e. using $t_0$, $u_0 w$, $t_{\rm eff}$, $t_{\star}$, 
$\pi_{\rm E,E}$,
$\pi_{\rm E,N}$,
$\alpha$,
$\gamma_{\parallel}$,
$\gamma_{\perp}$,
$s_z$,
$\log q$,
$\log w$, and
$\gamma_z$ as the MCMC parameters. These parameters are similar to those in Table \ref{tab:ulenspar}, with a few substitutions. In place of $t_\e$ and $\rho$ we have
$t_{\rm eff} \equiv u_0 t_\e$ and 
$t_{\star} \equiv \rho t_\e$. In addition, we use the caustic width $w$ instead of $s$ and $u_0 w$ instead of $u_0$ and step in the $\log$ of $q$ and $w$. As described in \citet{Skowron11}, this parameterization results in faster convergence of the MCMC.

In addition, we include the angular size of the source, $\theta_{\star}$, as a chain parameter. Although this is an observable quantity ($\theta_{\star,0} = 4.45\, \mu$as, see Section 4.1.1 of \citealt{Skowron11}), it has some uncertainty (7\%) for which we want to allow in the Markov chain. To do this, we allow this parameter to float, but we apply a $\chi^2$ penalty for values that deviate from the observed value, i.e.
\begin{equation}
\chi^2_{\rm penalty} = \left(\frac{\theta_{\star}-\theta_{\star, 0}}{\theta_{\star, 0}}\frac{1}{0.07}\right)^2.
\end{equation}
Finally, we change the sign of the RV data to account for the sign difference in the RV and microlensing coordinate systems \citep{Skowron11}.

For each link, the MCMC parameters are converted to binary orbit parameters, which are used to generate an RV curve. The degeneracy in $\pm(s_z, \gamma_z)$ leads to a degeneracy in the sign of the RV curve. In addition, the absolute RV offset, $\gamma$, is unmeasured. In order to determine the appropriate sign of the RV curve for each MCMC link, we generate the corresponding RV model and fit it to the RV data with both signs and optimize for the best value of $\gamma$ in each case. We take the better of the two fits as the ``correct'' sign, which determines the sign of $(s_z, \gamma_z)$ and also the values of $\Omega$ and $\omega_{\rm peri}$ (see Section \ref{sec:ulenspar}).

\begin{figure}
\includegraphics[height=0.75\textheight]{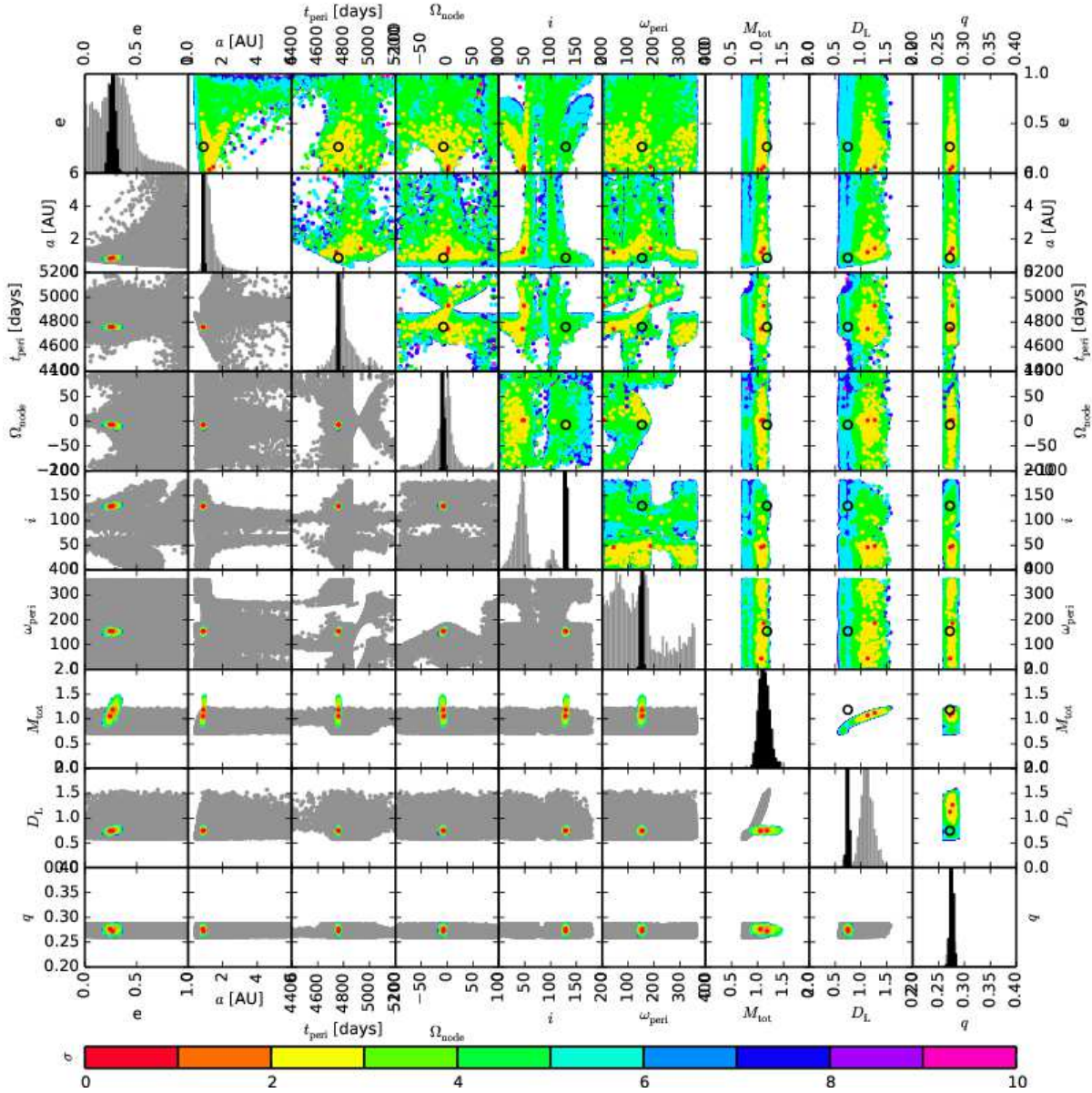}
\caption{Constraints on the orbit of \thisevent L. The upper right-hand panels show the constraints from the \citet{Skowron11} MCMC fit to the microlensing data. They are reproduced as the gray shaded regions in the lower left panels. The colored points in the lower left panels show the results of a joint MCMC to both the RV and
  microlensing data. The best joint fit is indicated by the black circle in the upper panels. The panels on the diagonal show the marginalized distributions of each parameter for the microlensing-only fit (gray) and the joint fit (black). \label{fig:jointconstraints}}
\end{figure}

The results of the MCMC are shown in the lower left panels of Figure \ref{fig:jointconstraints} in comparison to the constraints from \citet{Skowron11}. This clearly shows that including the RV data vastly improves the constraints on the orbital solution. To determine the final parameters of the system and their uncertainties, we weight the MCMC chain from the joint fit by the Jacobian \citep[Appendix B ][]{Skowron11} to account for the transformation from the MCMC parameters to physical parameters. The final values for the binary orbit are given in Table \ref{tab:orbittable}.

Note that \citet{Skowron11} require that the parameters of the lens star (flux, mass, and distance) are consistent with theoretical isochrones. This sets the upper and lower boundaries in $M_{\rm tot}$ for the microlensing-only chain. We do not include this weighting in our joint fits, which is why the posteriors extend to regions "excluded" by the microlensing MCMC.

\section{Conclusions\label{sec:conclude}}

We have performed the first test of a microlensing detection of lens
orbital motion by direct comparison of the microlensing orbit
constraints to the measured orbital parameters from radial velocity
observations of the lens system. Although the source and lens are not
resolved, we show that the ``contamination'' of the lens spectrum by
the source star is actually helpful. The fact that source is moving at
constant radial velocity allows it to serve as a wavelength reference
for our high-resolution spectra of the lens.  We find that the orbit
of {\thisevent}L as determined from radial velocity is fully
consistent with the constraints from the microlensing light curve,
which makes it the first confirmation of a microlensing measurement of
orbital motion.

Demonstrating that the parameters of the microlensing solution are consistent with RV follow-up is a very strong confirmation of
the method for including orbital motion in microlensing analysis. This test is completely independent of the microlensing data and is stronger than many previous tests of microlensing results because it constrains more parameters. Hence, we can now view the entire microlensing
orbital motion sample and the parameters we have derived with more confidence, including in the case of planetary microlensing events for which such followup is not possible.

In the future, a stronger test should be possible for the microlens
OGLE-2011-BLG-0417 \citep{Gould13_0417,Shin12}. While the
lens in this event is fainter, the microlensing constraints on the orbit are much better. In particular the form of the RV curve is predicted from the microlensing orbit measurement \citep[see Figure 1 of ][]{Gould13_0417}.

\acknowledgments

The authors would like to thank Josh Simon, Ian Czekala, Alicia Soderberg, and Atish Kamble for their assistance in obtaining RV data from Magellan.

Work by JCY was
performed under contract with the California Institute of Technology
(Caltech)/Jet Propulsion Laboratory (JPL) funded by NASA through the
Sagan Fellowship Program executed by the NASA Exoplanet Science
Institute.
AG was supported by NSF grant AST 1103471 and NASA grant NNX12AB99G. 
JSP was supported by a grant from the National Science Foundation Graduate Research Fellowship under grant no. DGE-1144469.
AV is supported by the National Science Foundation Graduate Research Fellowship, Grant No. DGE 1144152. 

This paper includes data gathered with the 6.5 meter Magellan Telescopes located at Las Campanas Observatory, Chile. Some of the data presented herein were obtained at the W.M. Keck Observatory, which is operated as a scientific partnership among the California Institute of Technology, the University of California and the National Aeronautics and Space Administration. The Observatory was made possible by the generous financial support of the W.M. Keck Foundation. The authors wish to recognize and acknowledge the very significant cultural role and reverence that the summit of Mauna Kea has always had within the indigenous Hawaiian community.  We are most fortunate to have the opportunity to conduct observations from this mountain.

{\it Facilities:} \facility{Keck:I (HIRES)}, \facility{Magellan:Clay (MIKE)}

\end{document}